\documentclass[11pt,twoside]{article}
\usepackage{asp2014}

\aspSuppressVolSlug
\resetcounters

\begin{document}

\title{What's the nature of sdA stars?}
\author{Ingrid Pelisoli$^1$, S.~O. Kepler$^1$, D. Koester$^2$ and A.~D. Romero$^1$}
\affil{$^1$Instituto de Física, Universidade Federal do Rio Grande do Sul, 91501-900 Porto~Alegre, RS, Brazil; \email{ingrid.pelisoli@ufrgs.br}}
\affil{$^2$Institut für Theoretische Physik und Astrophysik, Universität Kiel, 24098 Kiel, Germany}

\paperauthor{Ingrid~Pelisoli}{Author1Email@email.edu}{}{Universidade Federal do Rio Grande do Sul}{Instituto de Física}{Porto~Alegre}{Rio~Grande~do~Sul}{91501-900}{Brazil}
\paperauthor{S.~O.~Kepler}{kepler@if.ufrgs.br}{}{Universidade Federal do Rio Grande do Sul}{Instituto de Física}{Porto~Alegre}{Rio~Grande~do~Sul}{91501-900}{Brazil}
\paperauthor{D.~Koester}{koester@astrophysik.uni-kiel.de}{}{Institut für Theoretische Physik und Astrophysik}{Institut für Theoretische Physik und Astrophysik}{Kiel}{Schleswig-Holstein}{24098}{Germany}
\paperauthor{A.~D.~Romero}{alejandra.romero@ufrgs.br}{}{Universidade Federal do Rio Grande do Sul}{Instituto de Física}{Porto~Alegre}{Rio~Grande~do~Sul}{91501-900}{Brazil}

\begin{abstract}
White dwarfs with log~$g$ lower than 7.0 are called Extremely Low Mass white dwarfs (ELMs). They were first found as companions to pulsars, then to other white dwarfs and main sequence stars (The ELM Survey: 2010 to 2016), and can only be formed in interacting binaries in the age of the Universe. In our SDSS DR12 white dwarf catalog \citep{kepler2016}, we found a few thousand stars in the effective temperature and surface gravity ranges attributed to ELMs. We have called these objects sdAs, alluding to their narrow hydrogen line spectra showing sub-main sequence log~$g$. One possible explanation for the sdAs is that they are ELMs. Increasing the ELMs sample would help constrain the number of close binaries in the Galaxy. Interestingly, if they turn out to be A stars with an overestimated log~$g$, the distance modulus would put these young stars in the Galaxy's halo.
\end{abstract}

\section{Introduction}

White dwarf stars are the final evolutionary state of stars with initial masses up to 8.5--10.6~M$_{\sun}$ \citep{woosley2015}, corresponding to at least 95\% of all stars. For the evolution of single stars, the minimum mass of a white dwarf is around 0.30--0.45~M$_{\sun}${} \citep[e.g.][]{kilic07}, because progenitors that would become lower mass white dwarfs have main sequence evolution time larger than the age of the Universe. Such masses correspond, considering the mass-radius relation of white dwarfs, to a minimal log~$g$ of around 6.5. On the other hand, evolutionary models \citep[see][and references therein]{romero2015} indicate that the maximum log~$g$ of main sequence A stars, which have similar optical spectra to DA white dwarfs, is 4.75, even for very low metallicity.

Objects with $4.75<\log~g<6.50$ can result from binary evolution. Hot subdwarf stars are one example: binary interaction strips away the star's outer layers during core He burn, leaving a hot (T$_{\textrm{\tiny eff}}>$20\,000~K) lower mass (M$\sim$0.45~M$_{\sun}$) object. However, for low-mass progenitors (M$\lesssim$2.0~M$_{\sun}$), the temperature for burning He is only reached after it has become degenerate. Therefore, if the outer layers of a low-mass progenitor are stripped away before the He burning starts, a degenerate He core with a hydrogen atmosphere will be left: a white dwarf. Because the mass of the white dwarfs resulting from this channel can be much lower than the single star evolution limit (M$\lesssim$0.3~M$_{\sun}$), they are known as extremely-low mass white dwarfs, or ELMs \citep[see the ELM Survey:][]{ELMsurveyI, ELMsurveyII, ELMsurveyIII, ELMsurveyIV, ELMsurveyV, ELMsurveyVI, ELMsurveyVII}.

\section{Data Analysis}

Mining the Data Release 12 of the Sloan Digital Sky Survey \citep[SDSS DR12,][]{dr12} for white dwarfs, we found thousands of objects with hydrogen atmospheres showing $4.75 \leq \log~g \leq 6.5$. These objects were classified as type O, B, A, or white dwarf by the SDSS pipeline. Canonical mass white dwarfs were analyzed and published in \citet{kepler2016}. The remaining objects with signal-to-noise ratio larger than 15 were fitted to a grid of hydrogen-dominated atmosphere models, with metals added in solar abundances, covering 6\,000~K $\leq$ T$_\mathrm{\small eff}$ $\leq$ 40\,000~K and 3.5 $\leq$ log~$g$ $\leq$ 8.0. To choose between hot and cool solutions, we relied on the photometric results using the SDSS {\it ugriz} magnitudes and also the Galaxy Evolution Explorer \citep[\textsc{galex},][]{galex} {\it fuv} and {\it nuv} magnitudes, when available. This is the first time a fit with such large coverage in T$_\mathrm{\small eff}$ and log~$g$ is done to these spectra, to our knowledge.

This selection left 1010 objects with 8\,000~K $ \leq$ T$_\textrm{\tiny eff} \leq $ 20\,000~K and $5.00 \leq \log~g \leq 6.5$. About a hundred of them have proper motion higher than 10~mas/yr, most showing also galactic latitude larger than 30$^{\circ}$, indicating they may be outside the disk, since they are all fainter than $g$=14. As the nature of these objects is not yet fully understood, we have called them sdAs, referring to their narrow hydrogen line spectra (A-type) and their sub-main sequence $\log g$. If these objects are indeed main sequence A stars ($M_g\simeq 0$) with an overestimated log~$g$, some could be explained as relatively young high velocity or runaway stars. Only one high-velocity \citep{brown2009} and no runaway A stars are known to date. Moreover, their temperatures put them below the zero-age horizontal branch (ZAHB), so they cannot be explained as conventional He core burning subdwarfs, unless they are in binary systems as found by \citet{barlow2012}. In these sdB+FGK binary systems, the flux contribution of both components is similar, so the spectra appear to show only one object, with the lines of the main sequence star broadened by the presence of the subdwarf. The other possible explanation for these objects is that they are ELMs or pre-ELMs \citep{maxted2014a,maxted2014b}.

\section{Properties of sdA stars}

In an attempt to understand the nature of the sdAs, we have studied properties such as colors, proper motion, and velocities, comparing them to models and to the known ELMs listed in \citet{ELMsurveyVII}. Fig. \ref{ugr} shows the $(u-g)_0 \times (g-r)_0$ color-color diagram, where $u_0$, $g_0$ and $r_0$ are SDSS magnitudes with full extinction correction following \citet{schlegel1998}. The sdAs do not show the same colors as the known ELMs, but rather seem to extend the ELM branch to cooler temperatures. Evolutionary models \citep[e.g.][]{corsico2014,corsico2015} indicate that the time spent with 7\,500~K $ \leq$ T$_\textrm{\tiny eff} \leq $ 8\,500~K is about the same as the time spent with 8\,500~K $ \leq$ T$_\textrm{\tiny eff} \leq $ 22\,000~K. Less than 10\% of the known ELMs show T$_\textrm{\tiny eff} \leq 8500$~K \citep{ELMsurveyVII}, indicating that there's still a cool ELM population to be discovered, which is probably within the sdA sample. Curiously, despite the fact that the log~$g$ obtained from spectroscopy is above five, most sdAs lie below the model indicating log~$g = 5$ in this color-color diagram. This might indicate that spectroscopic log~$g$ is overestimated. Another possibility is that the extinction correction is not accurate for these objects.


\articlefigure[width=0.72\textwidth]{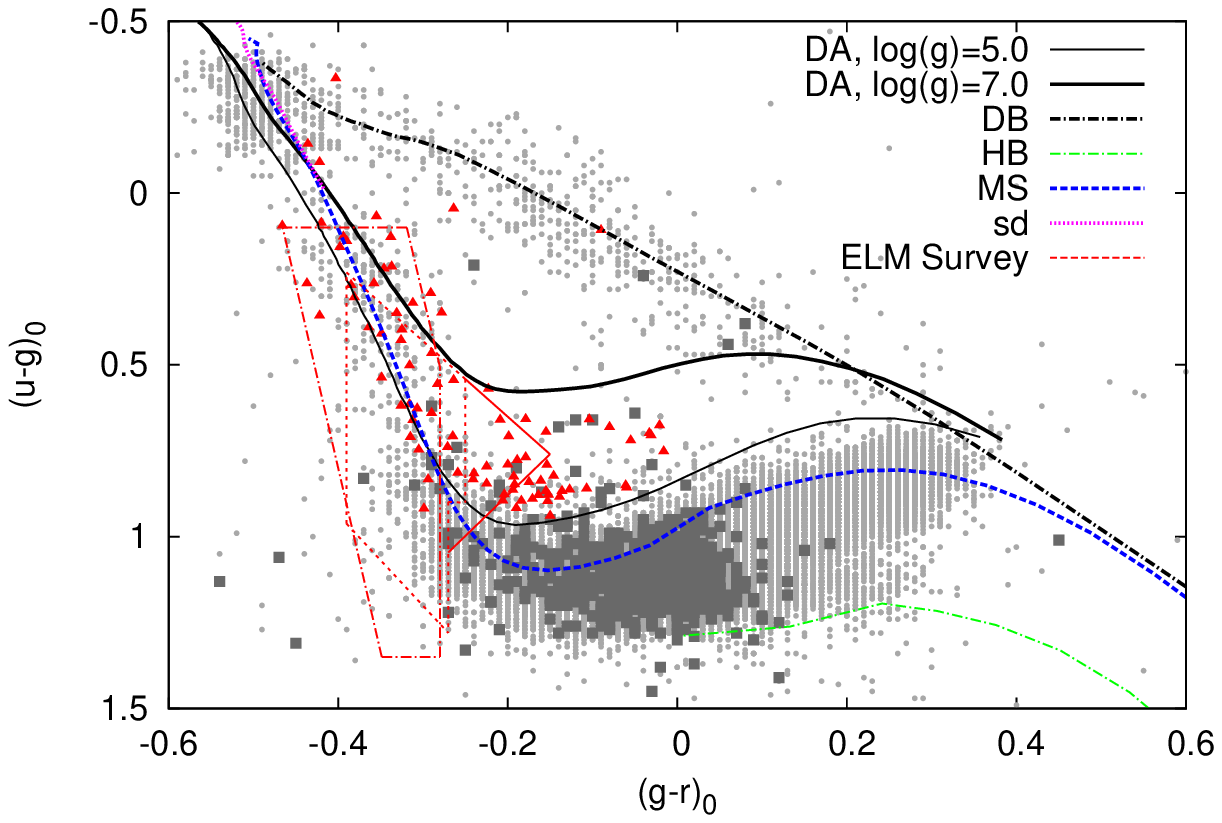}{ugr}{$(u-g)_0 \times (g-r)_0$ color-color diagram, showing the sdAs (squares, dark gray), the known ELMs (triangles, red), and the objects for which we obtained a $\log~g<5.0$ (dots, light gray) for comparison. The original color selection in the ELM Survey is shown in red, depending on magnitude: 15$ < g_0 < $17 dot-dashed line, 17$ < g_0 < $19.5 dashed line, and 19.5$ < g_0 < $20.5 continuous line. Theoretical models with fixed log~$g$ and metallicity are also plotted to guide the eye. Subdwarf (sd, magenta dotted line), main sequence (MS, blue dashed line), and horizontal branch (HB, green dotted-dashed line) models were calculated by \citet{lenz1998}. DB models (black dotted-dashed line) are from \citet{bergeron2011}. The DA models shown in black and labeled by log~$g$ were obtained from our grid of models.}

\articlefigure[width=0.72\textwidth]{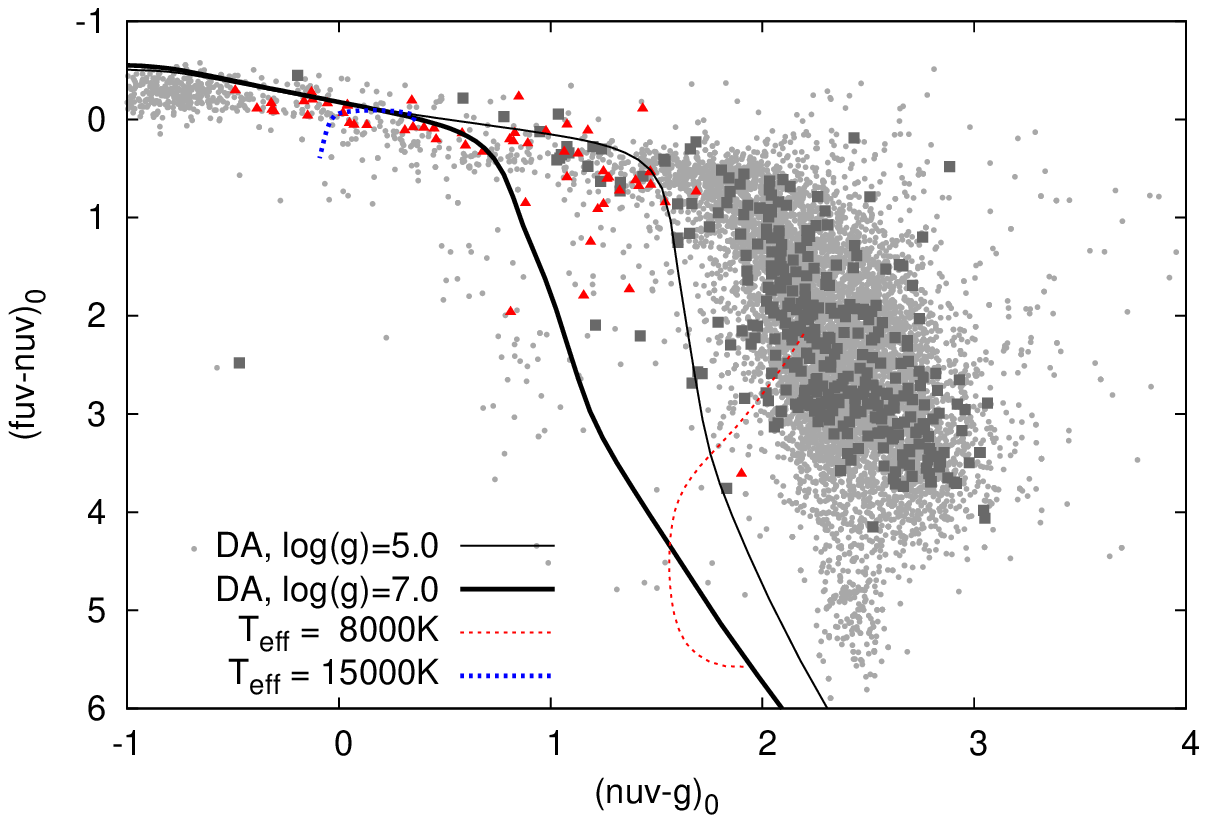}{uv}{GALEX colors the sdAs (squares, dark gray), the known ELMs (triangles, red), and the objects for which we obtained a $\log~g<5.0$ (dots, light gray) for comparison. Some DA models, both for fixed T$_\textrm{\tiny eff}$ and fixed log~$g$, are also shown on the plot.}

On Fig. \ref{uv}, a color-color diagram using \textsc{galex} colors for the objects available in this database is shown. Extinction correction was applied using the $E(B-V)$ given in the catalog, $R_{fuv}=8.24$ and $R_{nuv}=8.20$ \citep{wyder2007}. The picture here is the same as for the SDSS colors: the sdAs seem to be a cooler extension of the ELM branch. The fact that this stands for the UV colors is a strong suggestion that the sdB+FGK main sequence star hypothesis doesn't hold, because the sdB would have a strong contribution to the UV flux, which is not detected in most cases. 


Finally, by comparing the observed flux with the intensity given by the model, we obtained the observed solid angle, related to the ratio between the object's radius $R$ and its distance $d$. Assuming either a main sequence or an ELM radius for the sdAs, we estimated two possible values for $d$. Combining these distances with the measured proper motions obtained from USNO-B and SDSS data \citep{munn2004}, we estimated their galactic velocities U, V, and W \citep[e.g][]{johson1987} given the main sequence or the ELM radius. The results are shown on Fig. \ref{uvw}, together with the 3$\sigma$ ellipses for halo, thin, and thick disk as obtained by \citet{kordopatis2011}. When a main sequence radius is assumed, many objects show velocities considerably outside the halo distribution. Considering that main sequence A stars have typical lifetimes around 1--2~Gyr, they are expected to be found within the disk, so these velocities are highly unexpected. Some objects might be explained as relatively young high velocity (ejected by the Galaxy's central black hole) or runaway (ejected due to interaction in multiple systems) stars, but velocities higher than 1000~km/s cannot be explained by any of those mechanisms. Only one high-velocity \citep{brown2009} and no runaway A stars are known to date. The other possible explanation for these objects is that they are ELMs. When we assume an ELM radius, the objects have velocities consistent with a disk and halo distribution.


\articlefigure[width=0.7\textwidth]{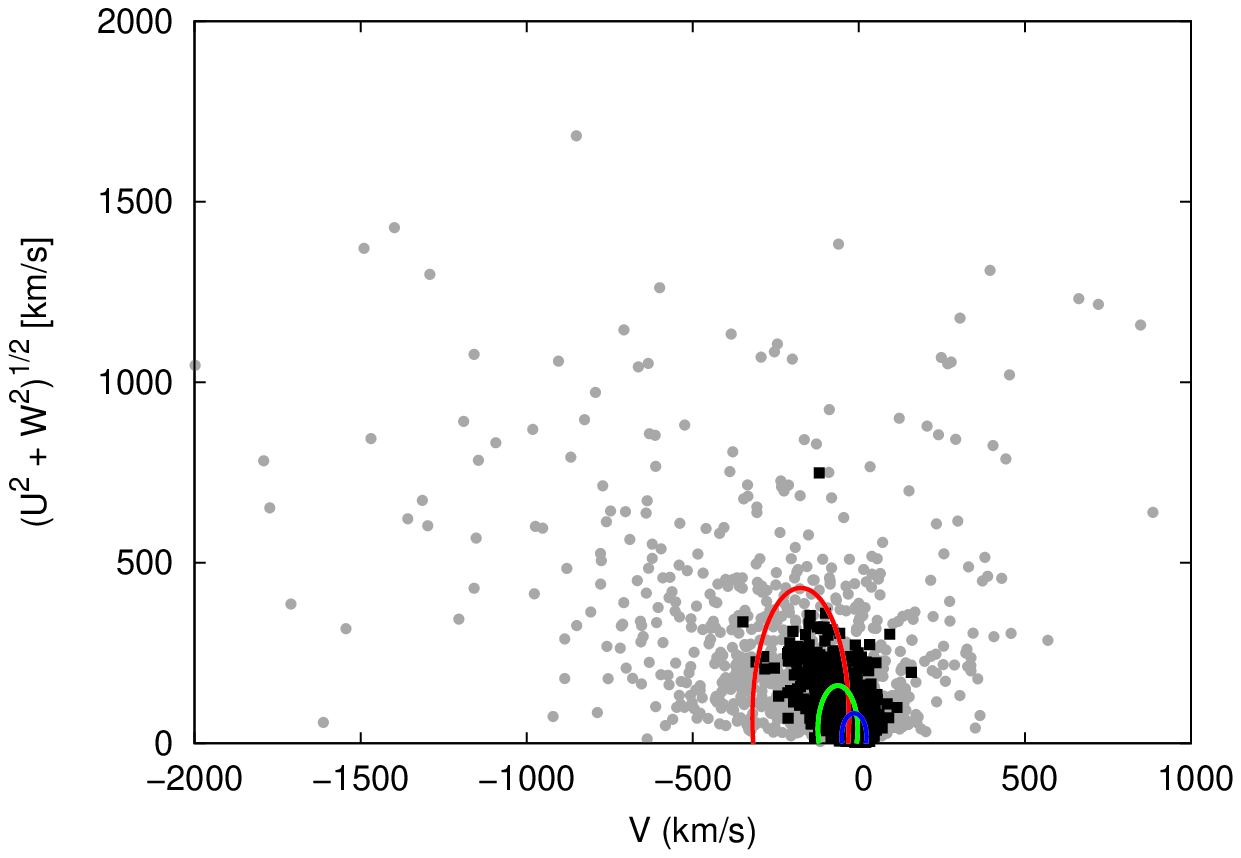}{uvw}{Estimated velocities for all sdAs assuming either a main sequence radius (dots, gray) or a ELM radius (squares, black). The ellipses show the 3~$\sigma$ limit for the halo (red), the thick disk (green), and the thin disk (blue), as given by \citet{kordopatis2011}.}

\section{Discussion}

There are three possible explanations for the nature of the sdAs: binaries of a subdwarf and a main sequence object of type FGK, main sequence A stars with an overestimated log~$g$, or ELMs. The first option is ruled out for most objects, because no significant flux in the ultraviolet, as would be expected from the subdwarf, is detected (Fig. \ref{uv}). Some objects do show UV flux [$(nuv-g)_0\leq1.5$, $(fuv-nuv)_0\sim0$], and thus might be explained as this kind of system. This can be verified with time-resolved spectroscopy to measure radial velocity variations. These systems have relatively long periods, of the order 3--4 years, showing radial velocity variations with a semi-amplitude typically smaller than 50~km/s \citep{barlow2012}. In contrast, ELMs, which are typically found in close binaries, have a median semi-amplitude of 220~km/s \citep{ELMsurveyVII}.

The main sequence A star hypothesis is strongly suggested by the colors shown by the sdAs. Both in the $(u-g)_0 \times (g-r)_0$ and the $(fuv-nuv)_0 \times (nuv-g)_0$ diagrams, the sdAs lie on cool regions, populated by low-log~$g$ objects, in spite of the log~$g > 5.0$ determined spectroscopically. As mentioned above, our models assume solar abundances for these objects, which is an overestimate considering the calcium abundances we have estimated from their spectra. Moreover, helium abundances were not taken into account. Finally, at the cool temperatures estimated for the sdAs, collisional effects between neutral hydrogen atoms should have a significant effect. There is no rigorous calculation of the line broadening caused by this effect in the literature yet, so it is also not properly accounted for. In short, there are many factors to be improved in the models, so we cannot rule out that the log~$g$ is not well determined. However, when we assume main sequence radii for the sdAs, the obtained distances and velocities are not consistent to what is expected for a population of young objects: they seem to be scattered through the halo, some with considerably high velocities which couldn't be explained even by black hole acceleration.

If the proper motion given in the SDSS tables are correct, hypothesis we tested by verifying they are similar to those given in the APOP catalog \citep{apop2015}, the only way to avoid having this scenario is if, instead of having a main sequence radius, these objects are actually ELMs, with a radius about twenty times smaller. This leads to a lower luminosity, in such a way that much smaller distances are needed to explain the observed flux. This also leads to much more reasonable velocities, within the thin and thick disk distributions, with very few objects appearing to be in the halo. The cooler colors presented by the sdAs when compared to the ELMs might be easily explained if they are a cooler ELM population, predicted by evolutionary models but sill underrepresented in the literature. The main issue with this hypothesis is that most sdAs do not show significant radial velocity variations between their SDSS subspectra. This might be explained if they are ELMs which already underwent a merging event. According to \citet{brown2016}, most double degenerate white dwarfs will merge within a Hubble time, so ELMs resulting from mergers are expected to exist. The known ELMs which shown no radial velocity are possibly explained as face-on systems. These two factors might explain the lack of variability to some objects, but is highly unlike that all the cool ELMs are either merged or in face-on systems. As the SDSS subspectra are usually of low-signal to noise and cover less than one hour, this can likely explain the no detection of variation to many objects. Therefore more observations are needed in order to understand the true nature of the sdA population.





\acknowledgements IP, SOK, and ADR are supported by CNPq-Brazil. DK received support from program Science without Borders, MCIT/MEC-Brazil. This research has made use of NASA's Astrophysics Data System and of the cross-match service provided by CDS, Strasbourg. Funding for the Sloan Digital Sky Survey IV has been provided by the Alfred P. Sloan Foundation, the U.S. Department of Energy Office of Science, and the Participating Institutions. SDSS-IV acknowledges support and resources from the Center for High-Performance Computing at the University of Utah. The SDSS web site is www.sdss.org.


\bibliography{Pelisoli_I}  


\end{document}